\documentclass[12pt,aps,prd,superscriptaddress,showpacs,longbibliography,floatfix,nofootinbib]{revtex4-2}

\usepackage[utf8]{inputenc}
\usepackage[english]{babel}
\usepackage [autostyle, english = american]{csquotes}

\MakeOuterQuote{"}
\usepackage{amsmath}
\usepackage{mathtools}
\usepackage{amsfonts}
\usepackage{physics}
\usepackage{amssymb}
\usepackage{MnSymbol} 
\usepackage{setspace}
\usepackage{graphicx}
\usepackage{xcolor}
\usepackage{hyperref}
\usepackage{slashed}
\usepackage{braket}
\usepackage{bbold}

\begin{document}
\title{Thermodynamic Susceptibilities for a Unitary Fermi Gas}
\author{Max Weiner}\affiliation{Department of Physics, University of Colorado, Boulder, Colorado 80309, USA}
\begin{abstract}
The unitary Fermi gas provides a unique window into both cold atom experiments and neutron star properties. There are major challenges in determining the physical properties within a neutron star, both experimentally and theoretically. However, there is a region within the crust of a neutron star that resembles a gas of fermions that is recreated in a cold atom laboratory. This is the so-called unitary Fermi gas characterized by a large negative scattering length and small effective range. We set out to calculate from first principles certain transport coefficients that appear at second order in the hydrodynamic expansion. These particular transport coefficients are obtained from two point correlation functions in flat space that do not require analytic continuation from Euclidean to Minkowski space. We are motivated by the potential to learn about neutron star properties from cold atom experiments.

\end{abstract}
\maketitle

\section{Introduction}
Currently, one of the prominent unresolved challenges in physics is understanding the properties of matter under conditions of ultra-high density and low temperature. Neutron stars (NS) are an embodiment of such a physical regime, yet the conditions within a NS are far removed from terrestrial experiments and observations of such objects are limited. Since their discovery in 1967 \cite{BellHewish1968} over 3,000 neutron stars have been cataloged, the vast majority detected as pulsars \cite{Manchester_2005}. This relatively small set of data, which mostly relies on a particular alignment of Earth with the axis of the NS beacon, puts into perspective the challenges experimentalists face. 

Theoretical obstacles are no less significant. Calculating properties of nuclear matter properties at high densities from first principles requires solving the equations of quantum chromodynamics (QCD), the accepted theory of nuclear interactions. Unfortunately, while efficient numerical methods exist for calculating QCD matter properties in equilibrium at small densities, these so-called lattice gauge theory calculations cannot be applied to high densities due to the infamous sign problem. In other words, the equations for calculating neutron star matter from first principles are known, but no one knows how to solve them.

The purpose of this paper is to calculate NS properties from first principles that can be measured in the lab. We are motivated by the fact that a region within the crust of a NS is composed of neutron-rich nuclei surrounded by a sea of free neutrons. Although free, these neutrons are strongly interacting, characterized by a large s-wave scattering length $a_s\approx -18.5$ fm and a small effective range $r_{\text{eff}}\approx 2.7$ fm \cite{Chen2008a_nn, Trotter1999a_nn}. When the scattering length is sent to infinity and the effective range to zero, all scales drop out and the system exhibits universal behavior since it is characterized only by its temperature, chemical potential, and density. Such a system is called a unitary Fermi gas (UFG) and the neutron sea we are interested in is close to this limit. Whether the fermions are atoms, electrons, quarks or something else, the way these systems interact in bulk will always be similar. UFGs have been studied extensively in cold atom laboratories where the s-wave scattering length of atomic fermions can be varied by applying an external magnetic field \cite{KetterleZwierlein2008}. Utilizing Feshbach resonances, the s-wave scattering length of certain atomic systems can be made very large such that effectively $k_F|a_s|\to \infty$, and sufficiently dilute such that $k_Fr_s\to0$, where the Fermi momentum is related to the density of the system via $k_F\sim n^{-1/3}$ \cite{Gurarie2007}. With this in mind, it may be possible to probe NS properties using table-top experiments. In particular, we are interested in transport properties of the UFG.

When a system is perturbed out of equilibrium, transport coefficients govern how the system relaxes back to equilibrium. Well-known examples include conductivities, viscosities, and diffusion constants \cite{LandauFluid_1959}. Over the past several years many experimental and theoretical studies have been devoted to hydrodynamic properties of strongly interacting quantum fluids, such as the UFG or the quark gluon plasma (QGP). It is interesting that these two systems are considered the most perfect fluids in the universe \cite{SchaferTeaney2009}, meaning their shear viscosity to entropy density ratio is close to the conjectured quantum lower bound $\eta/s\lesssim \frac{\hbar}{4\pi k_B}$ \cite{KSS_2005}. Transport coefficients such as the shear viscosity are, in general, difficult to calculate. However, in a remarkable paper Kovtun and Shukla \cite{KovtunShukla2018} showed that a certain family of transport coefficients, so-called thermodynamic transport coefficients, are considerably easier to calculate because they do not require analytic continuation from Euclidean to Minkowski spacetime. These quantities are classified as "thermodynamic" because they can be obtained by calculating certain thermodynamic susceptibilities in equilibrium.

Hydrodynamics is an effective low energy theory of systems near thermal equilibrium \cite{Romatschke2017relfld,KovtunHydro2012}. The expressions are valid on length scales much larger than the characteristic microscopic sizes of the system, such as the mean free path or correlation length. Hydrodynamics admits a derivative expansion in terms of the fluid variables such as temperature, chemical potential, and fluid velocity. At each order in this expansion are various transport coefficients quantifying the response of the system to external perturbations. Certain transport coefficients exhibit dissipation, leading to the generation of entropy in non-equilibrium conditions, while others remain non-dissipative and persist even in the limit of thermal equilibrium. Thermodynamic transport coefficients appear at second order in the gradient expansion and are non-dissipative in nature \cite{Baier_2008,Jensen_2012}.

These coefficients have been calculated for free uncharged relativistic scalar, vector, and fermionic quantum field theories \cite{RomatschkeSon_2009,MooreSohrabi,KovtunShukla2018} as well as free Dirac fermions at finite density \cite{ShuklaFermions}.  For interacting theories, results have been obtained for large $N$, $\mathcal{N}=4$ super Yang-Mills theory at infinite coupling for both zero and finite density utilizing the AdS/CFT conjecture \cite{Baier_2008,Bhattacharyya_2008,Finazzo_2015,GrozdanovStarinets_2015,GrieningerShukla_2021}. For coupling strengths between the weak and strong limits, analytic results were obtained in the large $N$ limit for the $O(N)$ model \cite{RomatschkeO(N)Kappa_2019,Weiner:2022vwa} and for cold non-relativistic fermions at finite density \cite{Lawrence:2022}. In this latter paper, Lawrence and Romatschke calculated the second order transport coefficient $\kappa$ for a UFG and we set out to calculate one more.

We work in natural units, $\hbar=c=k_B=1$.

\section{Calculation}
A system of fermions at unitarity begins with the Hamiltonian density in three spatial dimensions,
\begin{align}
	\mathcal{H} &= \sum_{s=\uparrow,\downarrow} \psi_s^{\dagger}(\textbf{x}) \left( \frac{-\grad^2}{2m} \right) \psi_s(\textbf{x}) + \frac{4\pi a_s}{m} \psi_{\uparrow}^{\dagger}(\textbf{x})\psi_{\downarrow}^{\dagger}(\textbf{x})\psi_{\downarrow}(\textbf{x})\psi_{\uparrow}(\textbf{x}),
\end{align}
where $\psi_s^{\dagger}$, $\psi_s$ are the fermionic creation and annhilation operators for states $s\in\{\uparrow$, $\downarrow\}$, respectively, $m$ is the fermion mass, and $a_s<0$ is the s-wave scattering length. Thermodynamic properties of this system are derived from its partition function at finite temperature, $T$ and chemical potential, $\mu$. The partition function in the path integral formulation is,
\begin{align}
	Z(T,\mu) &= \int \mathcal{D}\psi_s^{\dagger}\mathcal{D}\psi_s e^{-S_E}, \\
	S_E &= \int_0^{\beta} d\tau \int d^3\textbf{x} \left[ \psi_s^{\dagger}(\textbf{x}) \left(\partial_{\tau} - \frac{\grad^2}{2m} - \mu \right)\psi_s(\textbf{x}) + \frac{4\pi a_s}{m}\psi_{\uparrow}^{\dagger}(\textbf{x})\psi_{\downarrow}^{\dagger}(\textbf{x})\psi_{\downarrow}(\textbf{x})\psi_{\uparrow}(\textbf{x})\right], \nonumber
\end{align}
where $\beta\equiv 1/T$ is the inverse temperature and we utilize the Einstein summation convention to sum over the repeated indices $s$. We can make the path integral tractable by decoupling the four-fermi interaction with a Hubbard-Stratonovich transformation and introducing a complex auxiliary field $\zeta(\textbf{x})$,
\begin{align}
	\exp \left(-\int_{\tau,\textbf{x}} \frac{4\pi a_s}{m}\psi_{\uparrow}^{\dagger}\psi_{\downarrow}^{\dagger}\psi_{\downarrow}\psi_{\uparrow} \right) = \int \mathcal{D}\zeta  e^{\int_{\tau,\textbf{x}} \left(\frac{m}{4\pi a_s}\zeta\zeta^*-i\psi_{\downarrow}\psi_{\uparrow}\zeta^*+i\psi_{\uparrow}^{\dagger}\psi_{\downarrow}^{\dagger}\zeta \right)}
\end{align}
where we ignore an irrelevant normalization constant. Our partition function then reads,
\begin{align}
	Z(T,\mu) &= \int \mathcal{D}\psi_s^{\dagger}\mathcal{D}\psi_s \mathcal{D}\zeta e^{-S_{E,\text{eff}}} \label{eq:4} , \\
	S_{E,\text{eff}} &= \int_{\tau,\textbf{x}} \left[ \psi_s^{\dagger} \left(\partial_{\tau} - \frac{\grad^2}{2m} - \mu \right)\psi_s +i\psi_{\downarrow}\psi_{\uparrow}\zeta^*-i\psi_{\uparrow}^{\dagger}\psi_{\downarrow}^{\dagger}\zeta - \frac{m}{4\pi a_s}\zeta\zeta^* \right]. \nonumber
\end{align}
where all path integrals are now taken with respect to the \textit{effective} Euclidean action $S_{E,\text{eff}}$.

\subsection{Thermodynamics}
We may repackage the fermion bilinears into matrix form using Nambu-Gor'kov spinors, $\Psi = \left( \psi_{\uparrow},\psi_{\downarrow}^{\dagger}\right)$, such that the effective Euclidean action becomes,
\begin{align}
	S_{E,\text{eff}} &= \int_{\tau,\textbf{x}} \left[ \Psi^{\dagger} \left( \partial_{\tau} \mathbb{1} - \frac{\grad^2}{2m}\sigma_z - \mu\sigma_z + i\zeta^*\sigma_- - i\zeta\sigma_+ \right)\Psi - \frac{m\zeta\zeta^*}{4\pi a_s}\right],
\end{align}
where $\mathbb{1}$ is the $2\times2$ identity matrix, $\sigma_i$ denote Pauli matrices, and $\sigma_{\pm} \equiv \frac{1}{2}\left(\sigma_x \pm i\sigma_y\right)$. The path integral is now quadratic in fermion fields and can be integrated out exactly,
\begin{align}
	Z(T,\mu) &= \int \mathcal{D}\zeta e^{\int_{\tau,\textbf{x}}\left[\frac{m\zeta\zeta^* }{4\pi	a_s}+ \ln \left(-\det G^{-1}\right)\right]} 
\end{align}
where $G^{-1}(\zeta,\zeta^*)=\begin{psmallmatrix} \partial_{\tau}-\frac{\grad^2}{2m} - \mu & -i\zeta \\i\zeta^* &\partial_{\tau}+\frac{\grad^2}{2m} +\mu \end{psmallmatrix}$ and the minus sign in front of the determinant comes from the Jacobian $\int \mathcal{D}\psi_s^{\dagger}\mathcal{D}\psi_s=-\int\mathcal{D}\Psi^{\dagger}\mathcal{D}\Psi$.

Up to this point we have not made any approximations. To perform the path integral over the $\zeta$ fields we employ the non-perturbative resummation method introduced by Romatschke \cite{Romatschke2019resum1} by expanding the auxiliary field around the global zero mode, $\zeta(\tau,\textbf{x})=i\Delta + \zeta'(\tau,\textbf{x})$. In the $R0$ approximation (or equivalently the leading large $N$ expansion \cite{Veillette,PredragSachdev2007}) we neglect the field fluctuations $\zeta'$ and assume the zero mode $\Delta$ is real,
\begin{align}
	Z_{R0}(T,\mu) = \int_{-\infty}^{\infty}d\Delta e^{\beta V \left[ \frac{m\Delta^2}{4\pi a_s} + T\sum_{\omega_n}\int_{\textbf{p}} \ln\left(- \det \tilde{G}^{-1} \right) \right]}
\end{align}
where the sum is over fermionic Matsubara frequencies $\omega_n=2\pi T(n+1/2)$ with $n\in \mathbb{Z}$, $V$ is the volume of the system, and the inverse propagator in frequency-momentum space is
\begin{align}
	\tilde{G}^{-1}(\omega_n,\textbf{p},\Delta)=
		\begin{pmatrix} 
		i\omega_n+\epsilon_{\textbf{p}} - \mu & \Delta \\
		\Delta &i\omega_n-\epsilon_{\textbf{p}}+\mu 
		\end{pmatrix}\label{eq:8}
\end{align}
and $\epsilon_{\textbf{p}}\equiv\frac{\textbf{p}^2}{2m}$.

In the large volume limit, $Z_{R0}$ is solved exactly by the saddle point condition with respect to $\Delta$,
\begin{align}
	\frac{m\Delta}{4\pi a_s} + T\sum_{\omega_n}\int \frac{d^3\textbf{p}}{(2\pi)^3}\frac{\Delta}{\Delta^2+(\epsilon_{\textbf{p}}-\mu)^2+\omega_n^2}=0.
\end{align}
We see that there is the trivial solution $\Delta=0$, but we proceed to look for non-trivial solutions. After performing the sum over fermionic Matsubara frequencies we obtain the condition
\begin{align}
	\frac{-m}{4\pi a_s} = \frac{1}{2} \int \frac{ d^3\textbf{p}}{(2\pi)^3} \frac{ \tanh\left( \frac{\sqrt{\Delta^2 + (\epsilon_{\textbf{p}}-\mu)^2}}{2T}\right)}{\sqrt{\Delta^2 + (\epsilon_{\textbf{p}}-\mu)^2}}. \label{gap}
\end{align}
We note here that the poles of the Green's function, whose inverse is given in (\ref{eq:8}), define the spectrum located at $i\omega_n = \sqrt{\Delta^2+(\epsilon_{\textbf{p}}-\mu)^2}$. We see that excitations require a minimum energy input of $\Delta$ which characterizes a \textit{gapped} spectrum. We refer to (\ref{gap}) as the gap equation and such systems are characteristic of superfluids. 

At zero temperature the Matsubara sum becomes an integral, $T\sum_{\omega_n}\to \int \frac{d\omega}{2\pi}$, and the pressure is given by
\begin{align}
	p=\frac{\ln Z_{R0}}{\beta V}= \frac{m\Delta^2}{4\pi a_s} + \int \frac{d\omega d^3\textbf{p}}{(2\pi)^4} \ln \left( \Delta^2+(\epsilon_{\textbf{p}}-\mu)^2+\omega^2 \right).
\end{align}
The argument inside the the logarithm needs to be unitless so we regulate by subtracting $\int d\omega\ln\omega^2$. This ensures the integral over $\omega$ converges since in dimensional regularization $\int d\omega \ln \omega^2 =0$. Thus, in dimensional regularization the pressure is
\begin{align}
	p=\frac{m\Delta^2}{4\pi a_s} + \int \frac{ d^3\textbf{p}}{(2\pi)^3} \sqrt{\Delta^2+(\epsilon_{\textbf{p}}-\mu)^2}. \label{pres}
\end{align}
The integral over momenta is now divergent but this is taken care of using a well-known technique in dimensional regularization. We can expand the integrand in (\ref{pres}) as a binomial series
\begin{align}
	\sqrt{\Delta^2+(\epsilon_{\textbf{p}}-\mu)^2}=\sum_{n=0}^{\infty} \binom{\frac{1}{2}}{n} \frac{(-2\mu\epsilon_{\textbf{p}})^n}{\left(\Delta^2+\mu^2+\epsilon_{\textbf{p}}^2\right)^{n-\frac{1}{2}}} \label{eq:13}
\end{align}
and each integral is solved in dimensional regularization using \cite{Nishida},
\begin{align}
	\int \frac{d^D\textbf{p}}{(2\pi)^D}\frac{\epsilon_{\textbf{p}}^{\alpha}}{(\epsilon_{\textbf{p}}^2+A^2)^{\frac{\beta}{2}}} = \frac{\Gamma\left(\frac{D+2\alpha}{4}\right)\Gamma\left(\frac{2\beta-2\alpha-D}{4}\right)}{2\Gamma\left(\frac{D}{2}\right)\Gamma\left(\frac{\beta}{2}\right)}\left( \frac{mA}{2\pi}\right)^{\frac{D}{2}}A^{\alpha-\beta}. \label{dimreg}
\end{align}
We find that in the limit $D\to 3$,
\begin{align}
	\int \frac{ d^3\textbf{p}}{(2\pi)^3} \sqrt{\Delta^2+(\epsilon_{\textbf{p}}-\mu)^2}=\frac{2}{5}\frac{\mu(2m\mu)^{3/2}}{3\pi^2}g(y),
\end{align}
where $y\equiv \frac{\mu}{\sqrt{\Delta^2+\mu^2}}$, the function  is $g(y)=y^{-5/2}\left[(4y^2-3)E\left(\frac{1+y}{2}\right)+\frac{3+y-4y^2}{2}K\left(\frac{1+y}{2}\right)\right]$, and $E(x)$, $K(x)$, are the complete elliptic integrals of the first and second kind, respectively. At zero temperature, the non-trivial solution to (\ref{gap}) fulfills
\begin{align}
	2E\left(\frac{1+y}{2}\right)-K\left(\frac{1+y}{2}\right)=\frac{\pi y^{1/2}}{2\sqrt{2m\mu}a_s} \label{gapsln}
\end{align}
One can check that the non-trivial solution to (\ref{gap}) has lower free energy than the trivial solution $\Delta=0$ for all $a_s<0$. For the unitary fermi gas $a_s\to -\infty$ and the gap equation becomes $K\left(\frac{1+y}{2}\right)=2E\left(\frac{1+y}{2}\right)$ corresponding to the numerical soulution $y\approx 0.652$ or $\Delta \approx 1.162\mu$. It is noted in \cite{Lawrence:2022} that for small scattering lengths the solution is close to $y\approx 1$ and (\ref{gapsln}) can be expanded around this point to obtain the analytic result
\begin{align}
	\Delta = \mu \, e^{-\frac{m}{2\pi a_s \nu}-2+3\ln 2}, \quad |a_s|\ll 1 \label{anlyt}
\end{align}
Lawrence and Romatschke point out that this analytic solution is within ten percent of the numerical solution of (\ref{gapsln})  for all $a_s<0$. For the cold unitary Fermi gas,
\begin{align}
	\lim_{a_s\to -\infty}\Delta= \mu e^{-2+3\ln 2} \approx 1.083\mu \label{approx}
\end{align}
we see that the R0 approximation gives the correct order of magnitude: $\Delta/\mu =\mathcal{O}(1)$ \cite{Gandolfi2015}. At unitarity, $g(y)=y^{-3/2}E(\frac{1+y}{2})$ and so the pressure is given by,
\begin{align}
	p&=\frac{2}{5}\frac{\mu(2m\mu)^{3/2}}{3\pi^2}\frac{E(\frac{1+y}{2})}{y^{3/2}}\Bigg|_{y\equiv \left(1+\frac{\Delta^2}{\mu^2}\right)^{-1/2}}\approx 2.2 p_{\text{free}}
\end{align}
where $p_{\text{free}}=\frac{2}{5}\frac{\mu(2m\mu)^{\frac{3}{2}}}{3\pi^2}$ is the pressure of a cold \emph{non-interacting} fermi gas. The universality of a unitary fermi gas dictates $p=\xi^{-3/2}p_{\text{free}}$ where $\xi$ is a constant known as the Bertsch parameter. In our approximation we find $\xi_{R0}\approx0.59$ (which is what one obtains from a mean-field theory calculation \cite{Veillette}) which should be compared to the measured value of $\xi\approx0.38$ \cite{Ku_2012,Endres_2013}.

From the pressure we can obtain the number density,
\begin{align}
	\lim_{a_s\to-\infty}n(\mu)=\frac{\partial p}{\partial \mu}=\frac{(2m\mu)^{3/2}}{3\pi^2}\xi^{-3/2}
\end{align}
which is used to define the Fermi momentum
\begin{align}
	k_F\equiv (3\pi^2n)^{1/3}.
\end{align}

\subsection{Transport Coefficients}

Hydrodynamics provides an effective low energy description of physical systems near thermal equilibrium \cite{Romatschke2017relfld}. Hydrodynamics describes the dynamics of conserved macroscopic quantities of an underlying microscopic theory. The conserved hydrodynamic variable $T^{\mu\nu}$ (the energy-momentum tensor) is related to the symmetries of the microscopic theory via Noether's theorem. Further, we assume that these variables can be expressed as functions of a local temperature $T(x)$, a local chemical potential $\mu(x)$, and a local fluid four-velocity $u_{\mu}(x)$ normalized to $u^2=-1$.

The energy-momentum tensor can be decomposed in the following way \cite{KovtunHydro2012},
\begin{align}
	T^{\mu\nu}=\mathcal{E}u^{\mu}u^{\nu}+\mathcal{P}\Delta^{\mu\nu}+\mathcal{Q}^{\mu}u^{\nu}+\mathcal{Q}^{\nu}u^{\mu}+\mathcal{T}^{\mu\nu}
\end{align}
where $\Delta_{\mu\nu}=g_{\mu\nu}+u_{\mu}u_{\nu}$ is a projector transverse to the fluid four-velocity (e.g. $u^{\mu}\Delta_{\mu\nu}=0$), $\mathcal{E}=u_{\mu}u_{\nu}T^{\mu\nu}$ is the energy density, $\mathcal{P}=\frac{1}{3}\Delta_{\mu\nu}T^{\mu\nu}$ is the pressure, the energy flux $\mathcal{Q}^{\mu}=-\Delta^{\mu\alpha}T_{\alpha\beta}u^{\beta}$ is transverse to $u_{\mu}$, and the stress $\mathcal{T}^{\mu\nu}=T^{\langle \mu\nu\rangle}$ is transverse to $u_{\mu}$, symmetric, and traceless. The angled brackets denote the symmetric transverse traceless part of a tensor, $X_{\langle\mu\nu\rangle}=\frac{1}{2}\left(\Delta_{\mu\alpha}\Delta_{\nu\beta}+\Delta_{\mu\beta}\Delta_{\nu\alpha}-\frac{2}{3}\Delta_{\mu\nu}\Delta_{\alpha\beta}\right)X^{\alpha\beta}$.

Hydrodynamics assumes that the quantities $\mathcal{E}, \mathcal{P}, \mathcal{Q}^{\mu}$, and $\mathcal{T}^{\mu\nu}$ can be expanded in terms of the hydrodynamic variables $u^{\mu}$, $T$, and $\mu$ and their derivatives. These expressions are called the constitutive relations \cite{KovtunHydro2012}. Gradients of hydrodynamic variables correspond to perturbations out of equilibrium and each term is coupled to a quantity called a transport coefficient. Transport coefficients dictate the real-time evolution for equilibrium to be restored, familiar examples being conductivities, viscosities, and diffusion coefficients. These well-known quantities, however, control the linear (first-order) response of the system to a perturbation whereas non-linear contributions (second-order, third-order, etc.) will also be present. 

In general, transport coefficients of a strongly interacting quantum many-body system are difficult to calculate. However, for a particular subset of second-order transport coefficients, so-called thermodynamic transport coefficients, Kovtun \& Shukla showed such quantities are considerably easier to calculate from two-point correlation functions in flat space without the need to analytically continue from Euclidean to Minkowski spacetime \cite{KovtunShukla2018}. These transport coefficients are non-dissipative in nature and exist even in the limit of thermal equilibrium \cite{ShuklaFermions}. As first noted in \cite{Banerjee_2012,Jensen_2012}, these non-dissipative transport coefficients follow from an equilibrium generating functional which itself admits a derivative expansion, with various thermodynamic susceptibilities appearing as coefficients of different terms in the expansion. The variation of the equilibrium generating functional with respect to external sources produces the constitutive relations for the conserved currents in the system, which contain various transport coefficients that do not vanish in the equilibrium limit. Thus, these second-order thermodynamic transport coefficients are linear combinations of the thermodynamic susceptibilities. So we proceed to calculate susceptibilities and use the dictionary provided in \cite{KovtunShukla2018} to obtain transport coefficients. Susceptibilities are denoted as $f_i$, $i=\{1,2,3,...\}$ and the constitutive relations are,
\begin{subequations}
	\begin{align}
		\mathcal{E}&=\epsilon + \left(f_1'-f_1\right)R+\left(4f_1'+2f_1''-f_2-f_2'\right)a^2\nonumber\\
		&\quad+\left(f_1'-f_2-3f_3+f_3'\right)\Omega^2-2\left(f_1+f_1'-f_2\right)u^{\alpha}R_{\alpha\beta}u^{\beta},\\
		\mathcal{P}&=p+\tfrac{1}{3}f_1R-\tfrac{1}{3}\left(2f_1'+f_3\right)\Omega^2-\tfrac{1}{3}\left(2f_1'+4f_1''-f_2\right)a^2+\tfrac{2}{3}\left(2f_1'-f_1\right)u^{\alpha}R_{\alpha\beta}u^{\beta},\\
		\mathcal{Q}_{\mu}&=\left(f_1'+2f_3'\right)\epsilon_{\mu\lambda\rho\sigma}a^{\lambda}u^{\rho}\Omega^{\sigma}+\left(2f_1+4f_3\right)\Delta^{\rho}_{\mu}R_{\rho\sigma}u^{\sigma},\\
		\mathcal{T}^{\mu\nu}&=\left(4f_1'+2f_1''-2f_2\right)a_{\langle\mu}a_{\nu\rangle}-\tfrac{1}{2}\left(f_1'-4f_3\right)\Omega_{\langle\mu}\Omega_{\nu\rangle}+2f_1'u^{\alpha}R_{\alpha\langle\mu\nu\rangle\beta}u^{\beta}-2f_1R_{\langle\mu\nu\rangle},
	\end{align}\label{const}
\end{subequations}
where
\begin{align*}
	f_n'&\equiv Tf_{n,T}+\mu f_{n,\mu}\\
	f_n''&\equiv T^2f_{n,T,T}+2\mu Tf_{n,T,\mu}+\mu^2f_{n,\mu,\mu}
\end{align*}
and the comma denotes the partial derivative with respect to the argument that follows. Terms appearing in \ref{const} include the Riemann scalar $R$, Riemann tensor $R^{\mu\nu}$, acceleration $a^{\mu}\equiv u^{\lambda}\nabla_{\lambda}u^{\mu}$, and vorticity $\Omega^{\mu}\equiv \epsilon^{\mu\nu\alpha\beta}u_{\nu}\nabla_{\alpha}u_{\beta}$.\footnote{With the convention $\epsilon^{\mu\nu\alpha\beta}=\varepsilon^{\mu\nu\alpha\beta}/\sqrt{-g}$ and $\varepsilon^{0123}=1$.}

We set out to calculate $f_3$ given by \cite{KovtunShukla2018},
\begin{align}
	f_3&= \frac{1}{4}\lim_{\textbf{k}\to 0}\frac{\partial^2}{\partial k^2} \Big( -\langle T_E^{01}T_E^{01} \rangle (\textbf{k}) + \langle T_E^{12}T_E^{12} \rangle (\textbf{k})  \Big).
\end{align}
where $T_E^{\mu\nu}$ is the energy-momentum tensor of the underlying quantum system in Euclidean space. Without loss of generality we assume the external momentum is oriented in the $z$-direction, $\vec{k}=k\hat{z}$. We wish to calculate the Euclidean correlator
\begin{align}
	C_E(\omega,k\hat{\textbf{z}}) &\equiv  -\langle T_E^{01}T_E^{01} \rangle (\omega,k\hat{\textbf{z}}) + \langle T_E^{12}T_E^{12} \rangle (\omega,k\hat{\textbf{z}}).
\end{align}
In particular, we will calculate 
\begin{align}
	f_3 = \frac{1}{4}\lim_{k\to 0}\frac{\partial^2}{\partial k^2} C_E(\omega_n=0,k)\Bigg|_{k=0}
\end{align}

\subsection{Stress-Tensor Correlators in the Cold Superfluid Fermi Gas}
The Eucildean stress-tensor two-point correlator in the path integral formulation is
\begin{align}
	C_E(x-y) &= Z^{-1}\int \mathcal{D}\psi^{\dagger}\mathcal{D}\psi e^{-S_E} \left[-T_E^{01}(x)T_E^{01}(y) + T_E^{12}(x)T_E^{12}(y) \right].
\end{align}
We obtain $T_E^{\mu\nu}(x)$ by starting with the energy-momentum tensor of relativistic Dirac fermions and taking the non-relativistic limit (e.g. where the energy $m$ is much higher than any other energy scale in our system). We find
\begin{align}
	T_E^{01}(x) &= \frac{1}{4}\left[\psi_{\uparrow}^{\dagger}\left(\partial_1\psi_{\uparrow}\right) + \psi_{\downarrow}^{\dagger}\left(\partial_1\psi_{\downarrow}\right) - \left(\partial_1\psi_{\uparrow}^{\dagger}\right)\psi_{\uparrow} - \left(\partial_1\psi_{\downarrow}^{\dagger}\right)\psi_{\downarrow} \right] + \mathcal{O}(1/m) 
\end{align}
and $T_E^{12}(x) \propto \frac{1}{m}$, where we ignore all $\mathcal{O}(1/m)$ and higher terms. Expressing the stress-tensor component using Nambu-Gor'kov spinors, $\Psi=\begin{psmallmatrix}
\psi_{\uparrow}\\ \psi_{\downarrow}^{\dagger}
\end{psmallmatrix}$, our path integral becomes straight-forward to calculate. Since a vanishing total divergence of fields implies $\left(\partial_1\psi_s^{\dagger}\right)\psi_s=-\psi_s^{\dagger}\left(\partial_1\psi_s\right)$, $T^{01}_E$ simplifies to
\begin{align}
	T_E^{01}(x) = \frac{1}{2}\Psi^{\dagger}(x)\partial_1\Psi(x) + \mathcal{O}(1/m).
\end{align}
Using the same auxiliary fields as in (\ref{eq:4}) and to leading order in $1/m$, our path integral is
\begin{align}
	C_E(x-y) &= \frac{1}{4Z}\int \mathcal{D}\Psi^{\dagger}\mathcal{D}\Psi \mathcal{D}\zeta \Tr \left[ \partial_1^yG(y-x)\partial_1^xG(x-y) \right]
\end{align}
where the trace is over spinor indices and
\begin{align}
	\left<\Psi(x)\Psi^{\dagger}(y) \right> = G(x-y).
\end{align}
\indent In the R0 approximation we restrict the auxiliary fields to only the global zero mode. The R0-propagator from (\ref{eq:8}) is 
\begin{align}
	G(\omega_n,\textbf{p},\Delta)=\frac{1}{(\epsilon_{\textbf{p}}-\mu)^2+\omega_n^2+\Delta^2}
		\begin{pmatrix} 
			i\omega_n+\epsilon_{\textbf{p}} - \mu & \Delta \\
			\Delta &i\omega_n-\epsilon_{\textbf{p}}+\mu 
		\end{pmatrix}
\end{align}
and the stress-tensor correlator in frequency-momentum space $C_E(k_E,\textbf{k})$ at zero temperature is
\begin{align}
	C_E(k_E,\textbf{k}) = -\int_{-\infty}^{\infty} d\Delta \frac{e^{\beta V p}}{4Z} \int \frac{d^4p}{(2\pi)^4}p_1^2\Tr\left[G(k_E+\omega,\textbf{k}+\textbf{p})G(\omega,\textbf{p})\right]
\end{align}
where $\int d\Delta e^{\beta Vp}=Z_{R0}$ and the integral over $\Delta$ will once again restrict its value to the saddle point condition approximated in (\ref{approx}). Upon performing the trace over spinor components and restricting to zero external frequency $k_E=0$, our correlator is
\begin{align}
	C_E(k_E=0,\textbf{k})&=-\frac{1}{2} \int \frac{d\omega}{2\pi} \frac{d^3\textbf{p}}{(2\pi)^3}p_1^2\frac{(\epsilon_{\textbf{k}+\textbf{p}}-\mu)(\epsilon_{\textbf{p}}-\mu)+\Delta^2-\omega^2}{[(\epsilon_{\textbf{k}+\textbf{p}}-\mu)^2+\Delta^2+\omega^2][(\epsilon_{\textbf{p}}-\mu)^2+\Delta^2+\omega^2]} 
\end{align}
After $\lim_{\textbf{k}\to0}\frac{\partial^2}{\partial k^2}C_E$ and averaging over angles,
\begin{align}
	C_E''(K=0) &=-\int_P\left\{\tfrac{32 \epsilon_{\textbf{p}}^2 \left(\epsilon_{\textbf{p}}-\mu \right){}^4}{15 \left[\Delta ^2+\left(\epsilon_{\textbf{p}}-\mu \right){}^2+\omega ^2\right]^4}-\tfrac{4 \epsilon_{\textbf{p}} \left(13 \epsilon_{\textbf{p}}-5 \mu \right) \left(\epsilon_{\textbf{p}}-\mu \right){}^2}{15 \left[\Delta ^2+\left(\epsilon_{\textbf{p}}-\mu \right){}^2+\omega ^2\right]^3}+\tfrac{\epsilon_{\textbf{p}} \left(19 \epsilon_{\textbf{p}}-15 \mu \right)}{15 \left[\Delta ^2+\left(\epsilon_{\textbf{p}}-\mu \right){}^2+\omega ^2\right]^2} \right\}\nonumber\\
	&\quad-	 \Delta^2\int_P\left\{\tfrac{32 \epsilon_{\textbf{p}}^2 \left(\epsilon_{\textbf{p}}-\mu \right){}^2}{15 \left[\Delta ^2+\left(\epsilon_{\textbf{p}}-\mu \right){}^2+\omega ^2\right]^4}-\tfrac{4 \epsilon_{\textbf{p}} \left(7 \epsilon_{\textbf{p}}-5 \mu \right)}{15 \left[\Delta ^2+\left(\epsilon_{\textbf{p}}-\mu \right){}^2+\omega ^2\right]^3} \right\} \label{corr}\\
	&\equiv C_E^{a''}+C_E^{b''}
\end{align}
where the top line of (\ref{corr}) is $C_E^{a''}$ and the bottom line proportional to $\Delta^2$ is $C_E^{b''}$. Note that the integral in the bottom line of (\ref{corr}) is finite so that $C_E^{b''}$ vanishes in the limit $\Delta\to0$. We can rewrite the denominator of each term as

\begin{align}
	\frac{1}{\left[\Delta ^2+\left(\epsilon_{\textbf{p}}-\mu \right){}^2+\omega ^2\right]^{n}} = \frac{(-1)^{n-1}}{\left(n-1\right)!} \frac{\partial^{n-1}}{\partial(\Delta^2)^{n-1}} \frac{1}{\left[\Delta ^2+\left(\epsilon_{\textbf{p}}-\mu \right){}^2+\omega ^2\right]} 
\end{align}
for integers $n\geq1$ so that we are left with an integral over $\omega$ with simple poles at $\pm i\sqrt{\Delta ^2+\left(\epsilon_{\textbf{p}}-\mu \right)^2}$. The next step is to integrate over momenta utilizing equations (\ref{eq:13}) and (\ref{dimreg}) which, in the unitary limit $a_s\to -\infty$, leads to
\begin{align}
	C_{E}^{a''}(K=0)&=-\frac{(2m\mu)^{3/2}}{48\pi^2\mu}\frac{E\left(\frac{1+y}{2}\right)}{y^{3/2}}y^2 =- \frac{n(\mu)}{16\mu}y^2\label{alimit}\\
	C_{E}^{b''}(K=0)&=\frac{(2m\mu)^{3/2}}{720\pi^2\mu}\frac{E\left(\frac{1+y}{2}\right)}{y^{3/2}}\frac{8\mu^2+23\Delta^2}{\Delta^2}y^2 = \frac{n(\mu)}{16\mu}\frac{8\mu^2+23\Delta^2}{15\Delta^2}y^2 \\
	C''_E(K=0) &= \frac{n(\mu)}{30\mu} \left(\frac{\mu^2}{\Delta^2}\right)\\
	f_3&=\frac{1}{4}C''_E(K=0) = \frac{n(\mu)}{120\mu} \left(\frac{\mu^2}{\Delta^2}\right). \label{f3}
\end{align}
This is our final result for the thermodynamic susceptibility $f_3$. Second order thermodynamic transport coefficients are linear combinations of second order thermodynamic susceptibilities and their derivatives. These transport coefficients, as formulated in \cite{Baier_2008,Romatschke2017relfld}, and their relation to the susceptibilities are given in \cite{KovtunShukla2018}. For example, the second order transport coefficient $\lambda_3$ couples to the vorticity and, to leading order in the UFG, is given by $\lambda_3\approx -8f_3$.

\section{Conclusion}
The unitarity Fermi gas is an important example of a strongly correlated quantum fluid. Measurements of both equilibrium and non-equilibrium properties serves as a crucial reference point for a diverse range of physical systems, spanning from the study of dilute neutron matter in neutron stars to the investigation of the quark gluon plasma in relativistic heavy ion collisions.

Neutron star matter cannot be recreated in the lab and cannot be studied using lattice gauge theory because of the sign problem. Our result, equation (\ref{f3}), is an analytic solution from first principles and attempts to shed light on these issues. Because thermodynamic transport coefficients (which are linear combinations of thermodynamic susceptibilities as illustrated in equations (\ref{const}) do not require analytic continuation from Euclidean to Minkowski spacetime, our calculation is amenable to lattice calculations. Such calculations could further our understanding of such transport properties in NS crusts. Ultracold atomic experiments are known for their extraordinarily precise measurements. Since they are capable of creating UFGs in the lab, perhaps there is (or will be in the future) a means of measuring second-order transport coefficients. 

It has been suggested that superfluid vortices are relevant to NS glitches. Since $f_3$ couples to the vorticity (at second order) further study as to its relation to NS glitches might be worthwhile, although the relation between vorticity and superfluid vortices is unclear. 

Our results may be of assistance to those studying the NS equation of state (EoS). The authors of \cite{Tews2017ugc} propose a "unitary gas conjecture" which states that the energy of a UFG serves as a lower bound for pure neutron matter. This conjecture imposes constraints on EoS parameters.

There are six other thermodynamic susceptibilities that can be calculated for the unitary fermi gas \cite{ShuklaFermions}, one of which has already been calculated \cite{Lawrence:2022}. Calculating the rest of these transport coefficients would be the logical next step to this paper. The susceptibilities $f_1$, $f_2$, and $f_3$ characterize the constitutive relations in equations (\ref{const}), where $f_2$ has not been calculated for the UFG. The reason for this is that the two-point correlation functions used to determine $f_2$ consist of diagonal components of $T^{\mu\nu}$ which make it much more complicated. We leave this for future work.

\section{Acknowledgments}
This work was supported by the Department of Energy, DOE award No DE-SC0017905. I would like to thank Paul Romatschke for suggesting such an exciting and rich research topic as well as the enormous amount of help to make this work possible.

\appendix
\section{Free and Non-Relativistic Limit}
We compare our result in the free limit ($\Delta\to 0$) with the non-relativistic result in \cite{ShuklaFermions}. In that paper, Shukla calculated all thermodynamic transport coefficients for a relativistic system of free fermions. There he found,
\begin{align}
	f_3&=-\frac{|\mu_r|}{96\pi^2}\sqrt{\mu_r^2-m^2}
\end{align}
where we distinguish between the relativistic chemical potential $\mu_r$ and the non-relativistic chemical potential used in this paper, $\mu$. In the non-relativistic limit, the chemical potential is only slightly larger than the mass,
\begin{align}
	\mu_r\approx m+\mu
\end{align}
where $\mu\ll m$. We can then expand $f_3$ around $\mu=0$,
\begin{align}
	f_3&\approx -\frac{m^{3/2}\sqrt{\mu}}{48\sqrt{2}\pi^2} = -\frac{(2m\mu)^{3/2}}{192\pi^2\mu}= -\frac{n_{\text{free}}}{64\mu}
\end{align}
where $n_{\text{free}}=\frac{(2m\mu)^{3/2}}{3\pi^2}$ is the density of a degenerate Fermi gas. Looking at Eqns (\ref{alimit}) \& (\ref{f3}) and noting that $\lim_{\Delta\to 0}$ is equivalent to $\lim_{y\to1}$,
\begin{align}
	f_3=\frac{1}{4}\lim_{\Delta\to0} C_{E}^{a''}(K=0)&=\lim_{y\to1}-\frac{(2m\mu)^{3/2}}{192\pi^2\mu}\frac{E\left(\frac{1+y}{2}\right)}{y^{3/2}} = -\frac{(2m\mu)^{3/2}}{192\pi^2\mu} \label{limf3}
\end{align}
which agrees with the non-relativistic limit obtained in \cite{ShuklaFermions}. In order that our results agree with free fermions in the non-relativistic limit, we must also show that $\lim_{\Delta\to0} C_{E}^{b''}(K=0)=0$.

To this end, we must look to \cite{Gorda2022,Osterman2023} which conclude calculations (at zero temperature and finite chemical potential) that involve integrals over momenta and frequencies do not commute. That is, the physically correct results are obtained by integrating over momenta \textit{first}, and not vice versa as is commonly done. To this point we also note that when one sets $\Delta=0$ at the outset and goes on to calculate $\langle T^{01}T^{01}\rangle$ to leading order, integrating over frequencies first produces zero, which disagrees with \cite{ShuklaFermions}. It is straightforward from eqn (\ref{corr}) that $\lim_{\Delta\to0} C_{E}^{b''}(K=0)=0$ since it is the product of a finite integral and $\Delta^2$, so it must vanish in this limit.

\bibliography{ufg_f3}

\end{document}